\documentclass[12pt]{article}

\usepackage{amssymb, amsmath, amscd}
\usepackage{graphicx}
\usepackage[latin1]{inputenc}
\usepackage{natbib}
\usepackage{epsfig}
\usepackage{tikz-cd}
\usetikzlibrary{matrix,arrows,decorations.pathmorphing}
\usepackage{MnSymbol}
\usepackage{url}
\usepackage{lipsum}

\numberwithin{equation}{section}

\newcommand{\state}[1]{\left\vert #1 \right\rangle}
\newcommand{\dstate}[1]{\left\langle #1 \right\vert}

\newcommand{\Z}{{\mathbb Z}}

	{\begin{eqnarray} #1 \left\{ \begin{array}{lll}}%
	{\end{array} \right. \end{eqnarray}}

	{\begin{eqnarray}  \begin{array}{rcl}}%
	{\end{array}  \end{eqnarray}}

	{\\ \begin{tabular}{ll}}
	{\end{tabular}\\}

	{\begin{center} \hrulefill \quad {\sf #1} \quad \hrulefill \\[8pt]
		\begin{minipage}{0.90\linewidth}}
	{\end{minipage} \end{center} \hrule \vspace{2pt} \hrule}

   {\end{list}}

\begin{document}

\bibliographystyle{amsalpha}

\title{Very ample line bundles, contextuality and quantum computation}

\author{Raouf Dridi}
  \date{\today}
\maketitle

\begin{abstract} 
	I relate contextuality to line bundles.  Line bundles are important in algebraic geometry, they determine through their global sections  rational maps  to projective spaces. I  explain how such maps, if they exist,
 	relate rationally the input and output of measurement based computation (MBQC) and show geometrically that, indeed, contextuality is a necessary resource  for the computational advantage in MBQC.  I also leverage the definition of MBQC to category theory and present it as a ``subfunctor" of the spectral presheaf.  In general, 
	the MBQC functor is pointless whereas the computation is trivial.  	

\end{abstract}
% ----------------------------------------------------------------

\section{Introduction}
A line bundle is a rank one vector bundle, a familiar notion to theoretical physicists because of their occurrence in  geometric Berry's phase\footnote{Barry Simon, \emph{Holonomy, the quantum adiabatic theorem, and berry's phase},
  Phys. Rev. Lett. \textbf{51} (1983), 2167--2170.} for example. Here I explain how they also enter naturally into what we may call {\it contextual MBQC}.  
  Line bundles determine rational maps. I   explain how such maps
 relate input and output of MBQC and show geometrically that, indeed, contextuality is a necessary resource  for the computational advantage in MBQC. 
  
  \newpage
  
%\section{Contextuality through poset maps}
The convenient way for us to describe contextuality is using the {\it spectral presheaf}, denoted by $\Sigma$. Following \footnote{
A.~D{\"o}ring and C.~Isham, \emph{``{W}hat is a thing?'': topos theory in the
  foundations of physics}, New structures for physics, Lecture Notes in Phys.,
  vol. 813, Springer, Heidelberg, 2011, pp.~753--937. See also L. Loveridge, R. Dridi and R.  Raussendorf, \emph{Non-classical logic of classically universal measurement-based quantum computation}, http://arxiv.org/abs/1408.0745
}, one might think of $\Sigma$ as the generalized  phase space associated to the given quantum system. It is locally a set where locally here refers to the different windows (contexts) from which one has access/interacts to/with quantum mechanics.  I will present  $\Sigma$  as a poset\footnote{Partially ordered set.} map and mask the usual categorical definition; I also do the same, later on, with MBQC and presented as a poset map. Contextuality of the quantum system  translates to the following statement {\it the set of global sections (also called points) of $\Sigma$ is empty  (we also say $\Sigma$ is pointless)}\footnote{This leads to a second and more concrete motivation for the use of such mathematics.  
Indeed, from object-oriented programming persecutive (that is, if one decides to simulate contextual MBQCs using Python for instance), a  contextual MBQC is an instance of the class  {\sf pointless\_functor}  which should be completely deferent from the builtin class {\sf set}. }  \footnote{This also says that contextuality is not cohomological, set theoretical based structure suffices.}. A global section (a point) of $\Sigma$ is a  map between the two poset maps ${\bf 1}$ and $\Sigma$. 
 
 It is enough to go through the different constructions through the well-known example of Mermin's system\footnote{N.~D. Mermin, \emph{Simple unified form for the major no-hidden-variables
  theorems}, Phys. Rev. Lett. \textbf{65} (1990), no.~27, 3373--3376.}: 
  The Hilbert space
 is $\mathcal H = \mathbb{C}^{2^3}$ and the problem is to assign spectral values to the ten operators $$\sigma_x^1, \, \sigma_y^1,\, \sigma_x^2, \, \sigma_y^2,\, \sigma_x^3, \, \sigma_y^3, \, \sigma_x^1\otimes \sigma_x^2\otimes \sigma_x^3, \  \sigma_x^1\otimes \sigma_y^2\otimes \sigma_y^3, \, \sigma_y^1\otimes \sigma_x^2\otimes \sigma_y^3, \sigma_y^1\otimes \sigma_y^2\otimes \sigma_x^3.$$
 Such assignment is required to be compatible with evident constraints between commuting observables. 
 To prove this is not permissible using the spectral presheaf,  I proceed as follow:   Let $\mathcal A\subset \mathcal B(\mathcal H)$ be the non commutative 
C*-algebra generated by the ten different observables involved. I then define the poset $\mathcal W$ by its top layer given by the maximal abelian  subalgebras 
$$\mathcal W_1 :=  \{\sigma_x^1, \sigma_x^2, \sigma_x^3\}'', \, \cdots,
	 \mathcal W_5:= \{\sigma_x^1\otimes \sigma_y^2\otimes \sigma_y^3,\, \sigma_y^1\otimes \sigma_y^2\otimes \sigma_x^3,\,\sigma_y^1\otimes \sigma_x^2\otimes \sigma_y^3\}''.$$ 
By intersection, the maximal subalgebras $\{\mathcal W_i\}$ generate the poset $\mathcal W$.  The {spectral presheaf}  $\Sigma$ assigns
to each context $\mathcal W_i$ its Gelfan'd spectrum: 
$$\Sigma(\mathcal W_i) := \{\lambda_{\mathcal W_i}: {\mathcal W_i} \to \mathbb{C}~|~\lambda_{\mathcal W_i} \neq 0,~ \lambda_{\mathcal W_i}~ \text{is linear and } \lambda_{\mathcal W_i}(ab)=\lambda_{\mathcal W_i}(a)\lambda_{\mathcal W_i}(b)\}.$$   
To make this concrete, consider for instance  the last maximal context $\mathcal W_5$. The GHZ state  $GHZ := {|000\rangle + |111\rangle} $ is a common eigenstate for its operators. It defines an element in $\Sigma(\mathcal W_i)$  ({\it a local section of $\Sigma$})
by  $$\lambda_{GHZ} : \mathcal W_5 \rightarrow \mathbb C, \ A \mapsto \dstate{GHZ}A\state{GHZ}.$$ The remaining seven eigenstates define the remaining local sections of $\Sigma(\mathcal W_5)$. 

The second poset map {\bf 1} is the constant map which assigns to each subalgebra $W\in \mathcal W$, the singleton set $\{*\}$(any singleton set will do). A {global section} of the spectral presheaf is a map $\lambda: {\bf 1} \rightarrow \Sigma$  between the two poset maps ${\bf 1}$ and $\Sigma$. This translates into two conditions. Firstly, for all $W\in \mathcal W$ one has the map $\lambda_W:\{*\} \rightarrow \Sigma(W)$. What the map $\lambda_W$ does is to choose an element, $\lambda_W(*)$, from the set $\Sigma(W)$ (thus, the nature of the element inside the singleton set doesn't really matter!).  Note that, $\lambda_W(*)$ is a map itself i.e., a local valuation in $\Sigma(W)$.  Secondly, the map $\lambda$ is compatible with the poset structure. That is, for all $W'\subset W$, the restriction of the local valuation $\lambda_W(*)$ to $W'$ is the  local valuation $\lambda_{W'}(*)$. 

I  can now explain  how one can get the same inconsistent system for  Mermin's example. Assume that the spectral presheaf
posses a global section $\lambda: {\bf 1} \rightarrow \Sigma$.  For each maximal context $\mathcal W_i$, this global section will pick out a local valuation  $\lambda_{\mathcal W_i}(*)$. Taking into account the multiplicativity of  $\lambda_{\mathcal W_i}(*)$ and the fact
$\lambda$ satisfies the restriction condition above, one can safely drop the subscript $\mathcal W_i$ from   $\lambda_{\mathcal W_i}(*)$ and gets the familiar inconsistent system

 \begin{eqnarray}\nonumber
{\lambda}(*)(\sigma_x^1){\lambda}(*)(\sigma_x^2){\lambda}(*)(\sigma_x^3)=& {\lambda}(*)(\sigma_x^1\otimes \sigma_x^2\otimes \sigma_x^3) \\\nonumber
{\lambda}(*)(\sigma_x^1){\lambda}(*)(\sigma_y^2){\lambda}(*)(\sigma_y^3)=& {\lambda}(*)(\sigma_x^1\otimes \sigma_y^2\otimes \sigma_y^3) \\\nonumber
{\lambda}(*)(\sigma_y^1){\lambda}(*)(\sigma_x^2){\lambda}(*)(\sigma_y^3) = & {\lambda}(*)(\sigma_y^1\otimes \sigma_x^2\otimes \sigma_y^3) \nonumber \\\nonumber
{\lambda}(*)(\sigma_y^1){\lambda}(*)(\sigma_y^2){\lambda}(*)(\sigma_x^3) = & {\lambda}(*)(\sigma_y^1\otimes \sigma_y^2\otimes \sigma_x^3) \nonumber
\end{eqnarray} 
with $${\lambda}(*)(\sigma_x^1\otimes \sigma_x^2\otimes \sigma_x^3)  {\lambda}(*)(\sigma_x^1\otimes \sigma_y^2\otimes \sigma_y^3) {\lambda}(*)(\sigma_y^1\otimes \sigma_x^2\otimes \sigma_y^3) {\lambda}(*)(\sigma_y^1\otimes \sigma_y^2\otimes \sigma_x^3)=-1.$$ 

Similarly, the state dependent proof is obtained by requiring that the global section
restricts to a particular local section (for instance, restricts to $\lambda_{GHZ}$ on the fifth context). 

\section{Very ample line bundles}
 Line bundles are important in algebraic geometry \footnote{ For a very intuitive exposition about the basic facts used here, the reader might consider
the excellent {\it An Invitation to Algebraic Geometry} by K., Smith {\it et al.}}, they determine {\it rational}\footnote{
 % {\color{blue} Shafarevich page 35. }
A rational function on a variety $X$ is a function of the form $F(T_1, \cdots, T_n)/G(T_1, \cdots, T_n)$ 
such that $F(T_1, \cdots, T_n)$ and $G(T_1, \cdots, T_n)$ are polynomials and $G(T_1, \cdots, T_n)$ is  not a consequence of the defining equations of $X$.  A rational map $\varphi: X\rightarrow Y$ to a variety 
$Y\subset \mathbb P^m$ is an $m$-tuple of rational functions $\varphi_1, \cdots, \varphi_m$. The image of $X$
under a rational map $\varphi$ is the set of points $\varphi(X) = \{ \varphi(x)|\, x\in X \mbox{ and } \varphi \mbox { defined at } x\}$. } 
maps to projective spaces. Very ample line bundles determine {\it embeddings} in projective spaces.

Line bundles we need here are the {\it hyperplane} bundles.  The hyperplane bundle $H$ on a  variety $X$ is defined as follow: The fiber $\pi^{-1}(p)$ over $p\in X\subset \mathbb P^n$ is the 1-dimensional vector space of linear functionals on the line $\ell\in \mathbb P^n$ that determines $p$.  This line bundle has many global sections i.e.,  the homogenous coordinates on $X$.  
 
 Now, given a line bundle on $X$,  let us choose a set $\{s_0, \cdots, s_n\}$ of linearly independent global sections. These sections define the rational map
\begin{eqnarray}
 X &&\rightarrow \mathbb P^n, \\
 x&&\mapsto [s_0(x):\cdots:s_n(x)].
 \end{eqnarray} 
 The construction can be reversed: Every rational map $X \rightarrow \mathbb P^n$  is determined by some global sections of some line bundle over $X$. Indeed, the line bundle on $X$ will be the pullback of the hyperplane bundle on $\mathbb P^n$. 
 
   The vector space spanned by these sections is called
{complete linear system} if this vector space consists of all the global sections of the line bundle. A line bundle is called
{\it very ample} if the rational map determined by its complete linear system is an everywhere defined morphism that defines
an isomorphism onto its image\footnote{Let $X \rightarrow \mathbb P^n$ be the embedding of $X$ in projective space determined by a basis $s_0, \cdots, s_n$
of the global sections of a line bundle on $X$. Under this morphism, the sections $s_i$ become the coordinate functions. Thus, after embedding $X$ in $\mathbb P^n$ this way, the line bundle has become the hyperplane bundle. One may then think
of a very ample line bundle as one that, for some embedding of $X$ in projective space, is the hyperplane bundle.}.

\section{Spectral line bundle}\label{SpectralLineBundle}
In algebraic geometry, a vector bundle and its sheaf of sections are two equivalent data. 
%\footnote{Such connection wouldn't  be evident without the sheaf theoretic definition of contextuality}. 
Let us then identify each 
set $\Sigma(\mathcal W_i)$ (where $\mathcal W_i$ are the maximal contexts) with the hyperplane bundle $H_i\rightarrow X_i$ whose global sections are the local sections in $\Sigma(\mathcal W_i)$. The variety $X_i$ is defined by the different correlations between the local sections of  $\Sigma(\mathcal W_i)$ (for the $\ell_2$MBQC setting below, $X_i$ are linear vector spaces).  
 The local sections in $\Sigma(\mathcal W_i)$  represent now the coordinates functions on $X_i$ and their correlations represent the defining equations of $X_i$ expressed in this choice of coordinates. 
The spectral presheaf is a then locally, with respect to the cover $\{X_i\}$,  the hyperplane
$H_i\rightarrow X_i$.  %The different local sections of $\Sigma$ are local sections for
%the newly constructed line bundle\footnote {The spectral presheaf is the sheaf of sections of the line bundle. 
%{\color{blue} Also each global section of $\Sigma$   yields a global section for its line bundle. Indeed, if  $\lambda: {\bf 1} \rightarrow \Sigma$ then
%	$\mathrm{res}_{V_i} \lambda(*)$  are compatible local sections for the line bundle.  Inversely, if $s$ is a global section for the line bundlethen we have the natural transformation $\lambda: {\bf 1} \rightarrow \Sigma$ defined
%	by $\mathrm{res}_{V_i} \lambda(*) := \mathrm{res}_{V_i} s$.}
%	 }.
	 When this ``spectral line bundle'' is very ample, the spectral
	presheaf is a Gelfan'd spectrum. 

%
% 
%{\color{blue}
%NOT SURE about second condition in the def of vector bundle if it is satisfied here (when the intersection is more than one element). It should if
%the different masas are sent to each other through a linear transformation.  }
%\medskip
%
%{\color{blue}
%More details: For $\Sigma(V_1)$ we have the set of ps $U_1 := \{1, a, b, c, ab, ac, bc, abc \}$ and coordinates $\{s^j_1: U_1\rightarrow \Z_2\} = \Z_2^{U_1}$.
%For $\Sigma(V_2)$ we have the set of points $U_1 := \{1, a, b', c', ab', ac', b'c', ab'c '\}$ and coordinates $\{s^j_2: U_2\rightarrow \Z_2\}= \Z_2^{U_2}$.
%}

%The spectral presheaf is the sheaf of sections of $\pi$ (See section 8,3 page 123 Karen)

\medskip

%\begin{theorem}
%	$\lambda \pi = \lambda\Sigma$
%\end{theorem}
%\begin{proof}
%	
%\end{proof}
\section{Rational computations}

The following definition is from\footnote{\label{Robert} R. Raussendorf, \emph{Contextuality in measurement-based quantum
  computation}, Phys. Rev. A \textbf{88} (2013), 022322.}. An $\ell$-2 MBQC is  the following data:  a resource state $\varphi \in \mathbb{C}^{2^n}$, classical input ${\bf i} \in \mathbb Z_2 ^m$ and classical output $o : \mathbb Z_2 ^m \to \mathbb Z_2$, ${\bf i} \mapsto o({\bf i})$, a collection
of local observables $\{O_k({q_k}) |k \in \{1...n \}; q_k \in \mathbb Z_2 \}$ 
for which the measurement outcomes of a given $O_k({q_k})$ are labelled $s_k({q_k}) \in \{0,1\}$ for each $k$.
The computed output $o({\bf i})=\sum_{k} s_k({q_k})$ mod 2, and, with $q=(q_1...q_n),$ the measured observable $O_k({q_k}) $ is related to the outcome
$ s_k({q_k}) $ by $q=Q{\bf i}$ where $Q:\mathbb Z_2 ^m \to \mathbb Z_2 ^n$.

This definition mimics Mermin's example\footnote{Converted into $\ell$-2MBQC in J.~Anders, D.~Browne
 Computational Power of Correlations \emph{Phys. Rev. Lett}. {\bf 102}, 050502 (2009) } where the correlations between the local observables are multiplicative and 
which translates to linear correlations between the local sections. The argument presented below 
%(rational relationship between input and their correlations and output) 
is valid for more general  type of correlations and shows
the role of contextuality for the general scheme of MBQC: if the MBQC is non contextual then there is a rational map relating the different correlations (technically, coordinate rings) to the output i.e., the value of the global section. The importance of this particular instance of MBQC is that, once it is supplemented with a classical controller, one gets classical universal computation (if the MBQC is nonlinear).

In the language and notations of the previous sections, we have $n$ contexts $$\{\mathcal C_q := \{O_1(q_1),..., O_n(q_n)\}\}_{q = (q_1...q_n) \in \mathbb Z_2 ^n} $$ labled by $\mathbb Z_2 ^n$  plus a special context
 $$\mathcal C_{n+1} := \left\{O_1(q_1)\otimes...\otimes O_n(q_n) \right\}_{(q_1...q_n) \in \mathbb Z_2 ^m} $$ which stabilizes the resource state $\varphi$.    Each context yields a  maximal abelian subalgebras $\mathcal W_i := \mathcal C_i ''$.  The spectral presheaf
evaluated at $\mathcal C_q, \  q = (q_1...q_n)$ is the set $\{ v_k: \mathcal  C_q \rightarrow \Z_2, \  v_k({O_k({q_k}))} = s_k({q_k}), \ k=1..n\}$. The resource state defines a local section for the last context $\mathcal C_{n+1}$.  The computation
is quantum mechanical or contextual if the state dependent spectral presheaf  (where we require the global section to restrict to $\varphi$
on  the last context $\mathcal C_{n+1}$, in the same  we did with Mermin's system in Section 2) has no global section. 
We reproduce the result of ${}^{11}$: if the $l_2$MBQC is non contextual then the computed function is linear.  Indeed, the fact that
 the spectral presheaf $\Sigma$ posses a global section translates into a rational (in fact linear) map $\cup X_i\rightarrow \mathbb P^0$ i.e., to a point, {the output}, which represents the value of the section at the different observables. This rational map will also map linearly the different correlations (here linear correlations $o({\bf i})=\sum_{k} s_k({q_k})$) between the local sections for each context (in particular for the $n$  first contexts) to the coordinate values of the point. Thus, yielding a linear computation.

\section{Conclusion}
I have used the language of line bundles to describe the role of contextuality in MBQC.  
The key point was that, locally,  the spectral presheaf is the hyperplane bundle $H_i\rightarrow X_i$. Each variety $X_i$ is coordinated with the local
sections in $\Sigma(\mathcal W_i)$. Their correlations define the coordinate ring associated to $X_i$. 
 I have explained that if an MBQC, presented as a sub-functor of the spectral presheaf, is non contextual  then there is a linear map $\cup X_i\rightarrow \mathbb P^0$. This map relates the different correlations between the local sections (i.e., the coordinate rings of the varieties $X_i$) and in particular the first $n$  contexts representing the input   to the output i.e., the value of the global section.   %In short, contextual MBQCs are transcendental relations between input and output. 
 
To conclude, it would be interesting to see if this framework of functors and line bundles can actually be used in 
constructing examples of  what we may call {\it contextuality based computation}.
An {\it ideal} paradigm of quantum computation defined within this framework  which consumes contextuality whenever it is present and convert into a non trivial computation.

% We conclude with an interesting problem. Through the failure of the axiom choice, one can see that the intuitionistic logic of the underling topos is a {\it finer} necessary resource for the computation advantage for MBQC. The question is how one can make this explicit in the same way we did in this paper with contextuality. This might lead to {\it topos logic based quantum computation}.  

% The sets of local sections, $\Sigma(V_i)$
%for the first $m$ contexts are our input (or more precisely, a representative local section from each $\Sigma(V_i)$ will represent an input). 
%
%The fact that $s:X\rightarrow U_5$ is a global section translates as a rational function 
%between the input  (first $m$ contexts) and output which is here a point $\mathbb P^0$.  The global section $s$ is the resource state $\varphi$. 
%Indeed, the section will send the correlation inside the input contexts ( linear correlations $\sum_{k} s_k({q_k})$ in the case of $\ell$-2 MBQC) to 
%a given coordinate of the  point $\mathbb P^0$. 

%If $s$ is a global section then it defines a rational map $s:X\rightarrow U_5 =\{abc, ab'c', \cdots\}$ by $s(abc) = \mathrm{res}_{V_1} s(a) + \mathrm{res} _{V_1}s(b) + \mathrm{res}_{V_1} s(c) =: o(00) $ and $s(ab'c') =  \mathrm{res}_{V_2} s(a) + \mathrm{res}_{V_2} s(b') + \mathrm{res}_{V_2} s(c')= :o(01)   $ etc. 

%
%\bibliographystyle{plain}
%\bibliography{c}
% 
 
\end{document}